# COMPARISON OF COMMERCIAL THERMOLUMINESCENT READERS REGARDING HIGH-DOSE HIGH-TEMPERATURE MEASUREMENTS

P. Bilski[1], W. Gieszczyk[1], B. Obryk[1] and K. Hodyr[2]

[1]Institute of Nuclear Physics Polish Academy of Sciences, Radzikowskiego 152, Krakow, Poland
[2]Institute of Applied Radiation Chemistry, Lodz University of Technology, Wroblewskiego 15, Lodz, Poland

## ABSTRACT

Three different thermoluminescent measuring systems have been compared with respect to the differences in temperature profiles, spectral sensitivities, as well as characteristics for high intensities of TL light. The comparison was performed using the Harshaw 3500, Risø DA-20 and RA'94 TLD readers. The instruments were tested for the readouts of highly irradiated LiF:Mg,Cu,P (MCP) TL detectors, which require readout up to 550 $^{\circ}$C, in case of doses exceeding 1 kGy. It was found that the Harshaw 3500 can be used, without any additional light attenuation, for the measurements of MCP detectors exposed to doses up to about 5 Gy. For the other two readers the upper dose limit is about 5 times lower. It was also found that the Harshaw 3500 shows the best thermal stability considering the peak maximum position. For the ultra-high doses the differences in the spectral characteristics of the applied optical filters and photomultipliers, in conjunction with an evolution of the MCP TL emission spectrum with increasing dose, significantly influence the shape of TL glow curves measured with the DA-20 reader. The detailed characteristic of the compared TLD readers at high-dose high-temperature range is discussed.

# 1. INTRODUCTION

Typical, commercial thermoluminescent (TL) readers are usually designed to measure TL light intensity at the lowest possible level. In radiation protection, which is the most widespread application of thermoluminescence, the reason for this is a trend to obtain the lowest achievable dose detection limit. In TL dating and in material research this is caused by low sensitivity of the studied samples. However, there are situations, when there is a need to measure high intensity TL signals (e.g. for high doses in industrial dosimetry or even at lower doses, when an ultra-sensitive TL detectors are used). If readers are optimized for low light measurements, the question arises, what is their performance for high TL light intensities, close to the photomultiplier (PMT) detection limit? Some of readers offer a possibility to apply a neutral filter in order to avoid such conditions of work and to extend the measuring range. However, this is not always an acceptable solution, e.g. when in a series of measurements TLDs exposed both to low and high doses are present, so a wide dynamic range is needed.

Most of the standard TLD readers are also usually designed to perform measurements in the temperature range up to 400 $^{o}$C, what corresponds to the glow curves of common TL detectors. Moreover, the attention is mostly paid to the range below 300 $^{o}$C, where main dosimetric peaks typically occur. Nowadays, there are several commercial TLD readers with the extended temperature range, even up to 600 $^{o}$C, and there are applications, which really require measurements at such high temperatures. However, performance of these readers at the highest working temperatures has not been studied so far.

The aim of this work was to investigate and to compare three different TLD measurement systems with respect to high-dose and high-temperature measurements. Studies were realized using the RA'94, Harshaw 3500 and Risø DA-20 TLD readers. The motivation to undertake this work was the recently discovered thermoluminescence of LiF:Mg,Cu,P, at the high-dose and high-temperature range (Bilski *et al.*, 2008; Obryk *et al.*, 2009), and based on these effects the developed method of ultra–high dose measurements, in the kGy range (Obryk *et al.*, 2011a, Obryk, 2013). The shape of the LiF:Mg,Cu,P TL glow curve undergoes a complete alteration after irradiation above 1 kGy (a typical TL signal saturation level). These changes are related to the growth of high-temperature TL peaks, including so called peak "B", with maximum located at temperatures as high as 450 $^{o}$C or even higher (Obryk *et al.*, 2010, 2011b), what makes measurements even up to temperature of 550 $^{o}$C necessary. Furthermore, some significant changes in the LiF:Mg,Cu,P TL emission spectrum were also observed (Mandowska *et al.*, 2010; Gieszczyk *et al.*, 2013a, 2013b). These changes in the emission spectrum rise questions about influence of spectral sensitivity of readers on high-dose measurements, and this issue was also addressed in the work.

The investigations were realized using the LiF:Mg,Cu,P (MCP) detectors due to their high-sensitivity and presence of the above described high-temperature thermoluminescence. However, the observed effects and drawn conclusions are of a general kind and concern also other TL materials.

## 2. MATERIALS AND METHODS

*2.1. TLD readers specification*

Three different measurement devices, the Harshaw model 3500 (Thermo Scientific, USA), Risø model TL/OSL DA-20 (Risø DTU, Denmark) and RA'94 (Mikrolab, Poland) have been compared. All these readers are able to measure TL signal up to the temperature of 600 $^{o}$C or even 700 $^{o}$C (in case of the DA-20). The readers are equipped with photomultiplier tubes, metal resistance heaters and gas connections, which make it possible to measure TL signal in the neutral gas atmosphere. While the PMT tubes applied in the DA-20 and Harshaw 3500 are operated in photon counting mode, the one mounted in the RA'94 performs measurements in continuous current mode. Except for the differences in spectral sensitivities of different types of photomultipliers, the main difference between the studied devices is the heating method. In case of the Harshaw 3500 and RA'94 a sample is heated directly by a contact with the heating planchette. The DA-20 heats the sample indirectly via the stainless steel cup. This difference may cause some shifts of the measured TL glow curves towards higher temperatures. The another difference, important only from the practical point of view, is that the Harshaw 3500 and RA'94 are designed for a single detector measurements, while the DA-20 is equipped with the carousel, which allows to read out up to 48 detectors during one sequence. The readers are delivered with U340, BG39 and BG12 detection filters, in case of the DA-20 and RA'94, respectively. The standard detection filter, in case of the Harshaw 3500, is not specified by the producer. The reference light in the form of $^{14}$C isotope combined with a scintillator, in the RA'94, has been replaced by the LED diode in the other two readers. The most important characteristics of the studied devices are given in Table 1.

Table 1. The studied TLD readers specification.

|  | **HARSHAW 3500** | **RISØ TL/OSL-DA-20** | **RA'94** |
|---|---|---|---|
| Photomultiplier (PMT) | bialkali ETE 9125B | bialkali EMI 9235QB | bialkali EMI 9789QB |
| PMT mode | Photon counting | Photon counting | Continuous current |
| Standard filters | unspecified | U340, BG39 | BG12 |
| Heating method | contact, direct | contact, indirect | contact, direct |
| Heating range [$^{o}$C] | RT – 600 | RT – 700 | 40 – 600 |
| Number of samples in one sequence | 1 | up to 48 | 1 |
| Reference light | LED | LED | $^{14}$C + scintillator |
| Built-in irradiator | No | Yes | No |

*2.2. Irradiation and readout conditions*

All the TL measurements were performed using virgin LiF:Mg,Cu,P TL detectors, manufactured at the Institute of Nuclear Physics Polish Academy of Sciences (IFJ PAN), Krakow, Poland. The detectors in the form of sintered pellets of 4.5 mm diameter and 0.9 mm thickness were used. Before the irradiations a standard annealing procedure of 10 min at 240 $^{o}$C was applied. High–dose irradiations have been implemented at the Institute of Applied Radiation Chemistry, Lodz University of Technology, Lodz, Poland. The dose was delivered using the linear electron accelerator at the dose rate of about 1.7 kGy/min. Dosimetry was performed during the irradiations by the usage of radiochromic foils. The applied doses were ranging from 0.5 to 1215 kGy. Low–dose irradiations (doses below 0.5 kGy) have been realized at the IFJ PAN, Krakow using Sr-90/Y-90 beta source, mounted inside the DA-20, as well as an external Cs-137 gamma-ray source. The relative TL efficiency of MCP detectors, for all applied radiation qualities, is practically the same. Readouts were realized at the IFJ PAN, Krakow using three measurement systems, described in details in the previous section. In all cases the samples were heated up to temperature of 550 $^{o}$C with the constant heating rate of 2 $^{o}$C/s. MCP detectors after the readout up to 550 $^{o}$C were not longer used. In order to attenuate too high intensities of luminescence, at some of high–dose measurements, additional suppressing filters were applied. This was realized by the usage of the diaphragm with a hole of 0.5 mm diameter, in case of the DA-20 and RA'94. In case of the Harshaw 3500 the neutral density optical filter (Melles Griot ND-200) was utilized.

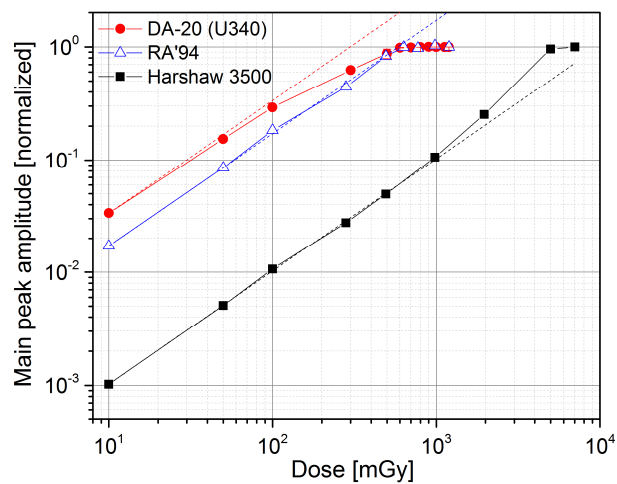

Fig. 1. Measurements of the detection systems' response in the range from the low doses up to the detection limit for the studied TLD readers.

## 3. RESULTS AND DISCUSSION

*3.1. PMT high-dose measuring limits*

Fig. 1 presents the results aimed on evaluation of the highest dose level that can be measured without any additional attenuation of the TL signal. The measurements were conducted under the standard HV values, recommended by the producers. It is visible that the Harshaw 3500 values are saturated by the signal of MCP TL detectors irradiated with doses around 5 Gy. This saturation dose value corresponds to the current per channel value of about 250 μA. In case of the DA-20 and RA'94 readers the dose limit is about 10 times lower. It was also found that the RA'94 shows the linear response up to the saturation level, while the response of the DA-20 and Harshaw 3500, close to the PMT measuring limit, becomes sub- and supralinear, respectively. It must be noted that this is not a result of the detector response characteristic, which is linear up to about 1 Gy and sublinear for higher doses. Such a response of the studied TLD readers may be caused by the fact that the RA'94 is operated in continuous current mode, unlike the other two readers, which are operated in photon counting mode. This finding suggests that a special attention is required while performing the readouts near the PMT measurement limit.

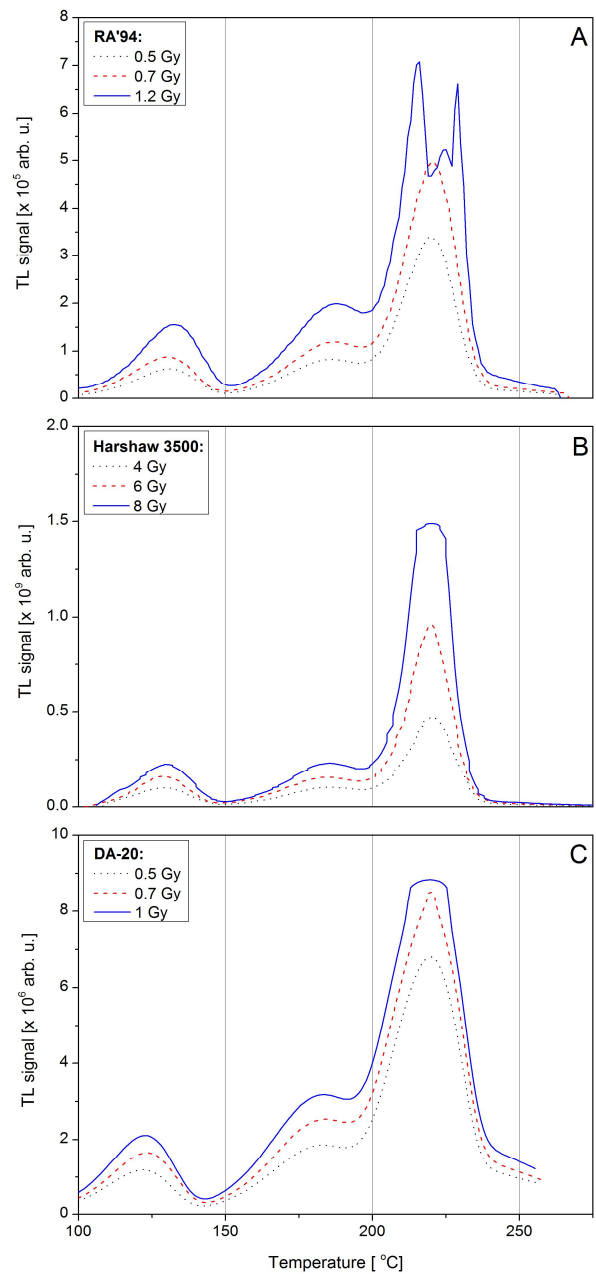

Fig. 2. Comparison of the MCP TL glow curves measured with the studied devices for signals exceeding PMTs detection limit.

Fig. 2 compares the TL glow curves measured with the studied devices for signals close and above the saturation threshold of the studied PMTs. It is visible that in case of the RA'94 (Fig. 2A) the shape of the main TL peak (around 220 $^{o}$C) practically does not change up to the detection limit. When the measured TL signal starts exceeding the saturation threshold, the top of the main TL peak becomes frayed. It is a unique property that was not observed for the other two readers. For the Harshaw 3500 and DA-20 (Figs. 2B and C, respectively), the glow-curves are deformed in another manner: the top of the main TL peak starts narrowing close to the measurement limit. At even higher values of the measured signal the top of the main TL peak becomes flat. These changes of the glow curve shape might be subtle and difficult to notice, unlike for the RA'94.

*3.2. Temperature stability and linearity*

The positions of various glow peaks were used to compare temperature calibration of the readers and to assess reproducibility of temperature control. Fig. 3 presents positions of the main dosimetric TL peak, at around 220 $^{o}$C, and the high-temperature peak, near 300 $^{o}$C, as the function of dose. The analyses were restricted to the dose ranges at which these peaks form apparent maxima (at least local), i.e. up to 10 kGy for the main peak and between 20 and 200 kGy for the high-temperature peak. For each radiation dose and each reader type 5 TLDs were processed. It should be noted that all data points plotted around a given dose, are actually related to this same dose value and they were just separated for better visibility. The spread of peaks position seems to be caused by not perfect contact between the heater and TLDs. It can be seen that the Harshaw 3500 shows the best thermal stability for both the main and high-temperature TL peaks. The other two readers present much worse characteristics. This can be explained by the differences in the heating systems. The Harshaw 3500 is equipped with more massive heater, as compared to the RA'94, that probably assures better thermal energy transfer to the sample. In turn, the DA-20 heats the sample indirectly via the stainless steel cup. This causes that the heating process may depend on various factors (e.g. surface impurities). The median temperature values for each TLD reader are marked in Fig. 3 by dashed lines. The difference in temperature calibration between the readers reaches 10 $^{o}$C for the main peak and nearly two times more for the high-temperature peak.

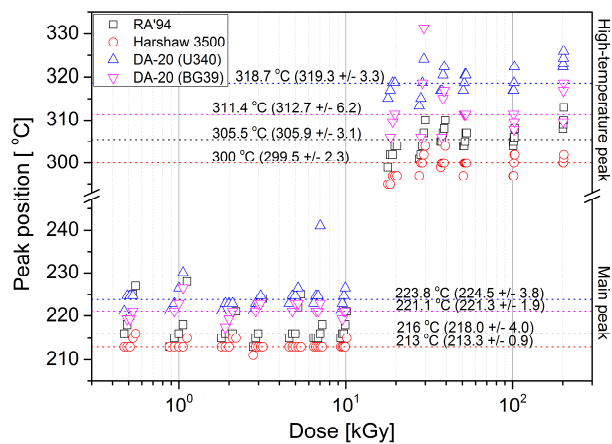

Fig. 3. Comparison of the thermal stability results for the studied TLD readers. The overlapping data points measured for a given dose were separated for better visibility.

In order to investigate higher temperature range the position of the ultra-high temperature peak "B" (see Fig. 5d) was exploited. The peculiar feature of this peak, which is visible for doses exceeding 40 kGy, is that its position shifts toward higher temperatures with increasing dose (Bilski *et al.*, 2008). In that way, each dose in this range produced a separate set of temperature related data.

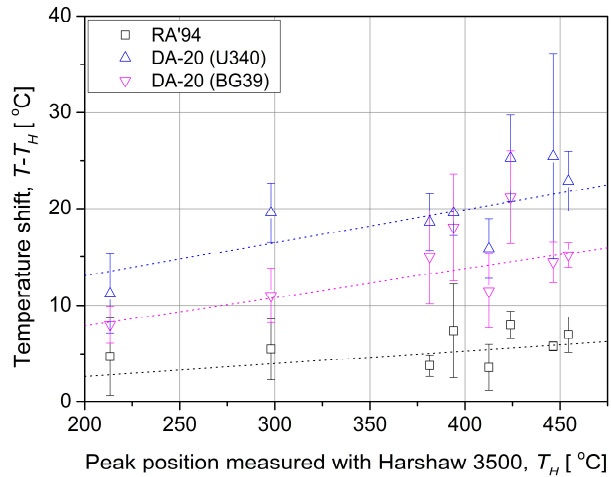

Fig. 4. TL peak temperature shift in reference to the values measured with the Harshaw 3500.

The peak positions measured with the Harshaw 3500 were treated as the reference values. This approach does not necessarily imply that these values are the most correct – they simply seem to be the most reliable, due to the lowest spread of the data. The difference between peak positions measured with a given reader with that of the Harshaw 3500 ($T-T_H$) are plotted in Fig. 4. From this figure it is visible that the temperature shift between the Harshaw 3500 and RA'94 is maintained well below 10 °C over the whole range of temperatures (a very small increase, observed for higher temperatures, seems to be within uncertainties). In case of the DA-20, the difference with the Harshaw 3500 depends on the temperature increasing from 10 °C, at low temperatures, up to 25 °C, near 450 °C. A somewhat surprising is the observed significant difference between values measured using the DA-20 with U340 and BG39 filters. A type of the filter has rather no influence on the TLD heating, so the explanation should be sought in the spectral effects, which will be discussed in the next section.

*3.3. Optical system spectral sensitivity vs. high-dose measurements*

Fig. 5 presents the MCP TL glow curves measured for different doses in the kilogray range. It is visible that at the "low" dose range, up to about 1 kGy (Fig. 5A), no significant differences are observed in the TL glow curves shape between the studied readers. Also for the highest dose of about 200 kGy, the glow curves shape is very similar for all the readers, with the exception of the temperature shift, mentioned in the previous section (Fig. 5D).

Completely different situation occurs for intermediate doses, close to 10 kGy (Figs. 5B and C), where each TLD reader produces somewhat different shape of a glow curve. The most apparent is the difference between glow curves measured with the DA-20 with two filters: the use of U340 causes a strong growth of the peak near 300 °C. As a result, for 9.74 kGy, this peak is the most prominent in the glow curve, unlike as for the other readers.

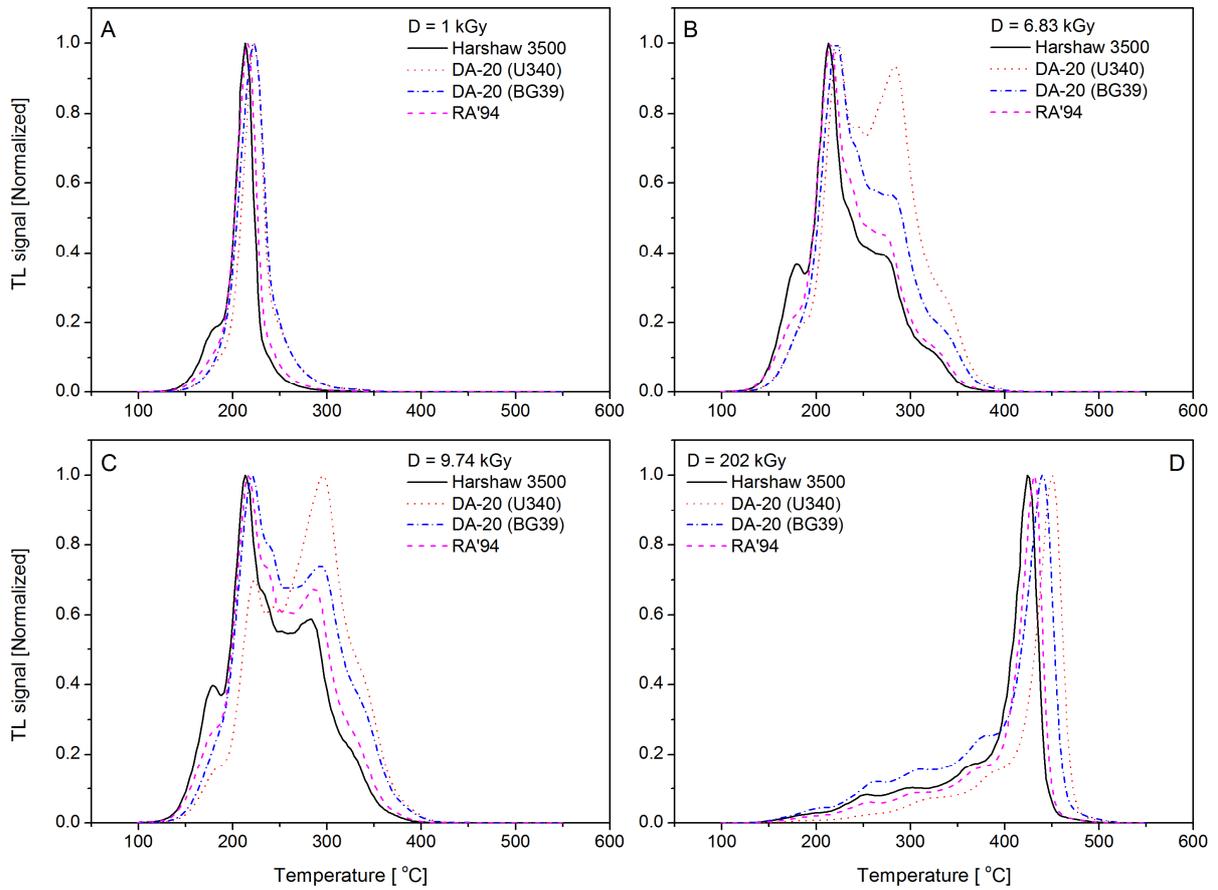

Fig. 5. Comparison of the MCP TL glow curves measured for the doses of 1 kGy (panel A), 6.83 kGy (panel B), 9.74 kGy (panel C) and 202 kGy (panel D) using the studied TLD readers.

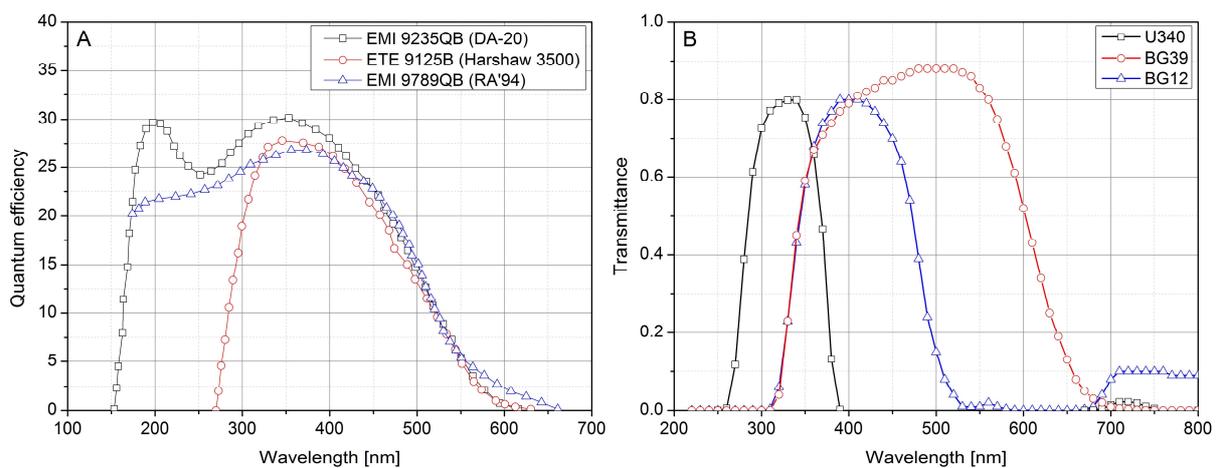

Fig. 6. Spectral characteristics of the applied photomultipliers (panel A) and optical filters (panel B).

This effect may be attributed to the mentioned in the Introduction changes of the emission spectrum. It was reported that an additional emission band in the MCP emission spectrum is observed at this dose range (Mandowska *et al.*, 2010; Gieszczyk *et al.*, 2013a, b). It seems that, at the high–dose range, the differences in the spectral characteristics of the applied optical filters and photomultipliers, in conjunction with an evolution of the MCP TL emission spectrum, significantly influence the shape of TL glow curves measured with the DA-20 reader. Spectral characteristics of the applied PMTs and optical filters are shown in Figs. 6A and B, respectively. One can see that spectral sensitivities of all studied optical filters and PMTs are comparable around 360 nm, where MCP TL emission maximum occurs. For the Harshaw 3500 the optical filter type is not specified by the producer, but obtained results tend to suggest that its characteristic is very similar to that applied in the RA'94. This thesis is also confirmed in Fig. 7 presenting the glow curve global maximum height as the function of dose. One can see that the values measured with the Harshaw 3500 and RA'94 are quite similar over a whole range of doses. At low doses the values measured with the DA-20 are also comparable. The highest differences are observed at the high–dose range, where the long wavelength emission band starts playing role in the MCP TL emission spectrum.

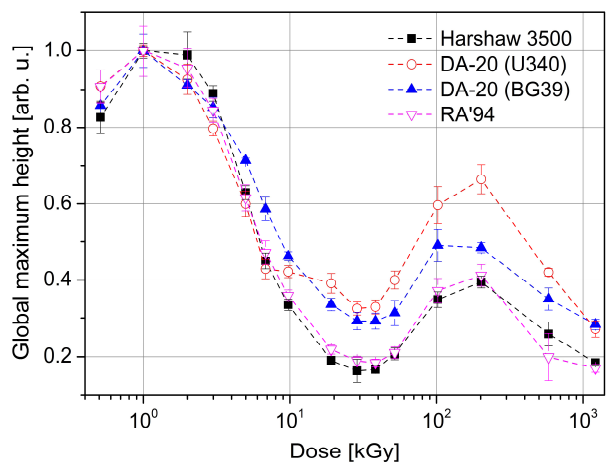

Fig. 7. Comparison of the global maximum height for the MCP TL glow curves measured with the studied TLD readers.

For the explanation of the observed differences two–dimensional emission spectrum, extracted from the previous measurements (Gieszczyk *et al.*, 2013b), measured for the dose of 8 kGy, is presented in Fig 8A. Fig. 8B compares the TL glow curves obtained as cross sections of Fig. 8A, at the wavelengths of 340 nm +/- 5 channels (maximum of the U340 filter) and 480 nm +/- 5 channels (maximum of the BG39 filter). It is clearly visible that the glow curves measured at the above mentioned ranges of wavelengths are almost identical to these presented in Fig. 5C. It fully confirms that differences observed in the measured TL glow curves shape result from different spectral characteristics of the studied devices.

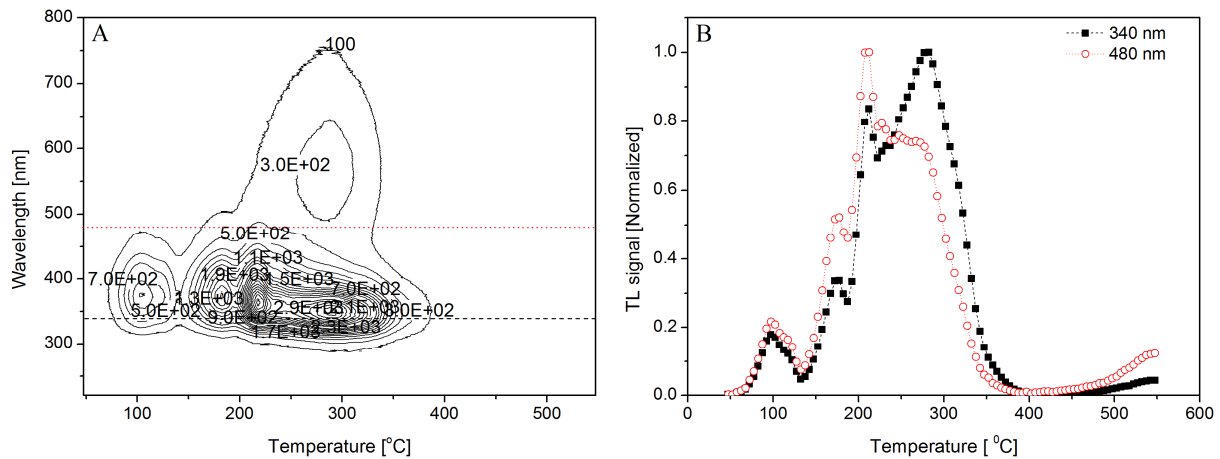

Fig. 8. Two-dimensional TL emission spectrum (panel A), measured for the MCP TL detector irradiated with the dose of 8 kGy. Panel B presents TL glow curves obtained as cross sections of the panel A, at the wavelengths corresponding to the maxima of the U340 (dashed line) and BG39 (dotted line) optical filters.

## 4. CONCLUDING REMARKS

Three different measurement systems, the Harshaw 3500, Risø DA-20 and RA'94 have been compared. Temperature profiles, spectral sensitivities and dose characteristics have been investigated. It was found that the Harshaw 3500 can be used to read out MCP TL detectors, irradiated with doses up to about 5 Gy (what corresponds to the current per channel value of about 250 µA) without any additional signal attenuation. For the other two readers this limit was about ten times lower. Each of the tested TLD readers exhibited different characteristic near the saturation level: the RA'94 was linear with dose, the DA-20 sublinear, while the Harshaw 3500 supralinear. The readers also show some deformations of the glow curve shape for the highest signals, again different for each TLD reader. These findings indicate that a special attention is required while performing the readouts near the PMT measurement limit.

From the analyzed devices the Harshaw 3500 shows the best thermal stability, probably due to its most massive heater. Temperature calibration (evaluated by positions of TL peaks) of the RA'94 differs from that of the Harshaw 3500 by 3-5 °C near 200°C and this difference does not change significantly over the whole studied temperature range. In case of the DA-20 this differences is about 10 °C at low temperature range and it doubles above 400°C.

The spectral sensitivity of the readers may significantly influence the measured high-dose response of TLDs. It was demonstrated for LiF:Mg,Cu,P, that even if various readers agree well at low doses, they differ considerably with respect to TL intensity, peak positions and glow curve shape at higher dose level. These effects are obviously a result of specific spectral characteristics of LiF:Mg,Cu,P, but one should suspect that such characteristics are not uncommon also for other materials.

# ACKNOWLEDGEMENTS

Work performed within the strategic research project "Technologies supporting the development of safe nuclear power" financed by the National Centre for Research and Development (NCBiR). Research Task „Research and development of techniques for the controlled thermonuclear fusion", Contract No. SP/J/2/143234/11.